\documentclass[aps,prd,preprint,superscriptaddress,nofootinbib]{revtex4-1}
\pdfoutput=1
\usepackage{graphicx}  
\usepackage{latexsym}
\usepackage{slashed}
\usepackage{dcolumn}   
\usepackage{bm}        
\usepackage{amsmath}
\usepackage{amssymb}   
\usepackage{longtable}
\usepackage{float}

\newcommand{\uah}{\affiliation{University of Alabama at Huntsville, Huntsville, AL 35805, USA}}

\begin{document}

\title{Inferred Hubble Parameter from Gravitational Waves in a Perturbative Bianchi I Background}

\author{Kevin J. Ludwick}
\email{kevin.ludwick@uah.edu} \uah
\author{Peter L. Williams}
\email{pw0029@uah.edu} \uah


\begin{abstract}
    It is straightforward to take the gravitational wave solution to first order in $v/c$ far from a binary source in a Minkowski background and adapt it to the Friedmann-Lemaitre-Robertson-Walker (FLRW) background, representing an expanding isotropic and homogeneous universe. We find the analogous solution for a slightly anisotropic background, which may be a more accurate description of our late universe through which gravitational waves propagate, and implications from tight CMB anisotropy constraints may not necessarily determine the level of anisotropy in the late universe in light of modified gravity models as well as the Hubble tension.  We use a perturbative form of the Bianchi I metric and demonstrate how the waveform differs. Using supernova anisotropy data as a reference, we show that the assumption of a Bianchi I background could imply on average a 2.1\% difference in inferred luminosity distance compared to what would be inferred under the assumption of the FLRW background.  This difference can be as high as 5.9\% depending on the observation direction.  Therefore, the background spacetime used for the inference of the Hubble parameter from gravitational wave data should be considered carefully.
\end{abstract}

\pacs{}\maketitle
\section{Introduction}

The advent of the detection of gravitational waves by LIGO has brought incredible new opportunities to our understanding of the cosmos.  It is well known that there is currently a Hubble tension, which is the disagreement of the estimate of the rate of the universe's expansion inferred from the cosmic microwave background (CMB) in the early universe and supernovae in the late, local universe. The determination of the Hubble parameter from gravitational wave (GW) data should give a more reliable value since it would be independent of the cosmic distance ladder, upon which supernovae analyses rely.
Therefore, it is important to infer the luminosity distance of GW sources accurately as they determine the inferred Hubble parameter.  A multiplicity of gravitational wave detections\cite{Ghosh, Pierra, LIGO2017} and multi-messenger detections\cite{Melo, LIGOmm} will afford a precise measurement of the Hubble parameter that is independent of the somewhat shaky rungs of the cosmic distance ladder\cite{Camarena}.

Obtaining the Hubble parameter from the CMB, supernovae, and GWs all assume the Friedmann-Lemaitre-Robertson-Walker (FLRW) metric and general relativity (GR) conventionally.  However, careful statistical analyses of supernova, quasar, and CMB data have suggested that an isotropic and homogeneous FLRW universe is not as accurate to a degree as usually assumed \cite{Secrest:2022uvx, Mohayaee:2021jzi, Secrest:2020has, Colin:2019opb, Cea:2022zep, Krishnan:2021jmh, Lombriser:2019ahl}.  For example, in \cite{Colin:2019opb}, the luminosity distance is expressed in a kinematic Taylor series to avoid making assumptions about the matter content or dynamics, and the inferred acceleration is essentially in the direction of the local peculiar bulk flow.  Other works that argue for negligible evidence for anisotropy \cite{Brout:2022vxf, Akarsu:2019pwn} typically assume an isotropic framework to some degree or make corrections for peculiar velocities by transforming to the ``CMB frame", which removes the evidence for anisotropy.  In \cite{Secrest:2020has}, the authors show that the distribution of distant matter as traced by radio sources and quasars is not isotropic in the CMB frame.  The authors of \cite{Secrest:2022uvx} demonstrate a $5.1\sigma$ rejection of the null hypothesis that there is a dipole anisotropy in the distribution of radio galaxies and quasars consistent with our inferred peculiar motion.  The question of bulk anisotropy may not be settled in the literature, but at the very least, metrics that account for anisotropy should be considered and studied.  

The CMB constraint on anisotropy is very stringent, constraining any shift away from isotropy in a particular direction of expansion to be $\lesssim 10^{-7}$ \cite{Schucker:2014wca}.  The Planck satellite data supports at most a very tiny amount of anisotropy \cite{Planck:2019evm}, and \cite{Hertzberg:2024uqy} show that any anisotropy in the early universe is washed out by expansion.  
However, the constraints coming from the early-time CMB need not agree with late-time
constraints from supernovae given the Hubble tension and growing support for MOND-like modified gravity models that can explain small scales well and may induce local anisotropies in structure formation, and since gravitational
waves propagate through a background that is late-time, only the supernovae
constraints were considered in this work. Another point to drive home is that the community is
eager to estimate a distance-ladder-free estimate of the Hubble parameter from
gravitational waves, but the inference of it may be incorrect due to the potential error in the
inference of luminosity distance, which this work endeavors to explore.

The accuracy of predictions of the waveforms of GWs due to massive binary systems has been refined to a high level over the years through advancements in numerical relativity. In the duration between emission of these waves from the source and detection by a gravitational wave (GW) observatory, the waveform is to a large degree unaffected by the spacetime through which it propagates. However, as more detections of GWs are made and more observatories come online with higher precision, our ability to discern subtle effects on the waveforms will increase.  Many works have studied gravitational waves, including their exact solutions in some cases, in background spacetimes other than FLRW, such as Bianchi-type backgrounds \cite{11, 12, 13}.  Others have examined the effect of propagation through cosmic dust and in the presence of electromagnetic fields and scalar fields \cite{21, 22, 23}, and much analysis on gravitational waves has been done in modified gravity models \cite{31, 32, 33}.  Our goal in this work is different in that we specifically want a solution that matches with the parameters of the source binary system without having to resort to an intensive numerical solution while also getting a solution easily comparable to the FLRW background case.  We want this in order to make the point that one should be cautious in inferring luminosity distance while taking the usual FLRW background for granted.

In this work, using supernova anisotropy data as a reference, we show that the assumption of a Bianchi I background spacetime could imply an average of 2.1\% difference (up to 5.9\% depending on the direction of observation) in inferred luminosity distance compared to what would be inferred under the assumption of the FLRW background.  The effect on the waveform is primarily in the amplitude as opposed to a more subtle effect on the oscillation pattern. 

\section{Bianchi I Metric}

The simplest possible expanding universe that obeys the cosmological principle (homogeneous, isotropic) is described by the FLRW metric,   

\begin{equation}
    d\tau^2=dt^2 - a(t)^2 (dx^2 + dy^2 + dz^2),
\end{equation}
where $a(t)$ is the scale factor which represents the changing scale of the universe.
Note that the speed of light $c$ has been set to $1$. A fairly intuitive way to generalize this is to drop the condition of isotropy, allowing the scale factor to vary independently in each coordinate direction. The result is the Bianchi I metric, described as 
\begin{equation}
\label{Bianchimetric}
    d\tau^2=dt^2-a(t)^2dx^2-b(t)^2dy^2-c(t)^2dz^2.
\end{equation}

where $a$, $b$ and $c$ are arbitrary functions of time.  
We have chosen to work with a perturbative form of this metric for the ease of comparing the waveform to the solution with an FLRW background, where $b$ and $c$ differ from $a$ only slightly:  

\begin{align}
\label{scales}
    &b(t)=a(t)(1+\beta), \\ \nonumber
    &c(t)=a(t)(1+\gamma),
\end{align}

where $\beta \ll 1$ and $\gamma \ll 1$.  In what follows, we keep to first order in $\beta$ and $\gamma$.  In general, $\beta$ and $\gamma$ are functions of time, but we also assume here that $\beta$ and $\gamma$ are slowly varying: $\dot \beta,  \dot \gamma \ll1$. 

\subsection{Gravitational Waves}

There are analytic solutions available for Bianchi space in general and for GWs for the region outside the binary source in certain cases \cite{Mondal:2020iet, Hu:1978td, Datta:2022ibj, Valent:2022smp}, but we want to obtain a solution in the region outside the binary source in terms of the binary source in order to have a waveform in terms of the luminosity distance from the source.  Creating full waveform solutions would be very time-consuming, so we consider the ``local wave" zone approach to account for the effect of propagating through a Bianchi I background.  The ``local wave" zone approach involves correcting the Newtonian solution for the strain to include backreaction due to the background metric in a ``local wave" zone that is far enough from the source so that the GWs have the right inverse source-observer distance dependence but far enough so that the universe's expansion is negligible during propagation in this region, as is done in chapter 4 of \cite{Maggiore}.  The authors of \cite{Pandey:2022kol} calculate the difference in luminosity distance in a (non-Bianchi) viscous cosmology and show how the inference of the Hubble parameter would change due to this difference.  In a similar fashion, we take into account the difference in luminosity distance and find the change in the GW waveform itself due to a Bianchi I background metric as it differs from the FLRW case.  

The perturbation with respect to the Bianchi I background, the strain, is defined as $h_{\mu \nu}$ by
\begin{equation}
g_{\mu \nu} = \gamma_{\mu \nu} + h_{\mu \nu},
\end{equation}
where $\gamma_{\mu \nu}$ is the background Bianchi I metric and $g_{\mu \nu}$ is the full metric.


To obtain the wave equation, we solve Einstein's equation to first order as outlined in \cite{Mondal:2020iet}.  The Riemann tensor is written as
\begin{equation}
R^\rho_{\mu \nu \lambda} = \bar{R}^\rho_{\mu \nu \lambda} + R^{(1)\rho}_{\mu \nu \lambda},
\end{equation}
and the Einstein tensor is written as 
\begin{equation}
G_{\alpha \beta} = \bar{G}_{\alpha \beta} + G^{(1)}_{\alpha \beta}= 8 \pi (\bar{T}_{\alpha \beta} + T^{(1)}_{\alpha \beta}).
\end{equation}
Since the perturbation to the stress-energy tensor $T^{(1)}_{\alpha \beta}$ is primarily a long-wavelength correction, it is decoupled from the short-wavelength regime and irrelevant for the evolution of gravitational waves of observational interest, as discussed in \cite{Maggiore}.  We use the Lorenz gauge
\begin{equation}
\nabla_\mu \tilde{h}^{\mu \nu} = 0,
\end{equation}
where $\tilde{h}_{\mu \nu} \equiv h_{\mu \nu} - \frac{1}{2} \gamma_{\mu \nu} h$, and $h = h^\alpha_\alpha$, the trace of $h_{\mu \nu}$.  In this case, Einstein's equation at first order is $G^{(1)}_{\alpha \beta} = 0$, given by Eq. (14) of \cite{Mondal:2020iet}:
\begin{equation}
\label{Eineq}
\square \tilde{h}_{\beta \alpha}  -2\bar{R}_{\delta \beta \alpha \rho} \tilde{h}^{\delta \rho} - \bar{R}_{\delta \alpha} \tilde{h}^\delta_\beta - \bar{R}_{\delta \beta} \tilde{h}^\delta_\alpha =0,
\end{equation}
where $\square \equiv \nabla^\alpha \nabla_\alpha$.

Without loss of generality, we consider a wave propagating in the $x$ direction.  We also use the transverse-traceless (TT) gauge, in which $\tilde{h}_{\mu \nu} = h_{\mu \nu}$.  The propagation in the $x$ direction, the transverse condition, the traceless condition, and the symmetry of the metric tensor imply the following conditions:
\begin{equation}
h_{0 \alpha}=0,~
h_{1 \alpha}=0,~
h_{22}=-h_{33},~
h_{23}=h_{32}.
\end{equation}
Then $h_{\mu \nu}$ is
\begin{equation}
\begin{pmatrix}
0 & 0 & 0 & 0\\
0 & 0 & 0 & 0\\
0 & 0 & h_{22} & h_{23}\\
0 & 0 & -h_{23} & h_{22}.
\end{pmatrix}
\end{equation}
As a result, Eq. (\ref{Eineq}) gives two non-zero independent equations.  Using Eq. (\ref{Eineq}), $(G^{(1)}_{22}-G^{(1)}_{33})/2$=0 gives
\begin{equation}
\label{h22}
-\left(\frac{\dot{b}^2}{b^2}+\frac{b \dot{b} \dot{c}}{c^3} +\frac{c \dot{b} \dot{c}}{b^3}+\frac{\dot{c}^2}{c^2}\right)h_{22} + \left(\frac{1}{a^2}\frac{\partial^2}{\partial x^2}+\frac{1}{b^2}\frac{\partial^2}{\partial y^2}+\frac{1}{c^2}\frac{\partial^2}{\partial z^2}\right) h_{22} +\left(-\frac{\dot a}{a} + \frac{\dot b}{b} + \frac{\dot c}{c}\right)\dot h_{22} - \ddot h_{22}=0, 
\end{equation}
where $\dot{}$ denotes differentiation with respect to $t$, and $(G^{(1)}_{23}+G^{(1)}_{32})/2$=0 gives
\begin{equation}
\label{h23}
-\frac{4\dot{b} \dot{c}}{bc} h_{23} + \left(\frac{1}{a^2}\frac{\partial^2}{\partial x^2}+\frac{1}{b^2}\frac{\partial^2}{\partial y^2}+\frac{1}{c^2}\frac{\partial^2}{\partial z^2}\right) h_{23} +\left(-\frac{\dot a}{a} + \frac{\dot b}{b} + \frac{\dot c}{c}\right)\dot h_{23} - \ddot h_{23}=0.
\end{equation}

Under the ``local wave approximation" (which allowed us to use $G^{(1)}_{\mu \nu}=0$), the background dynamics are slow compared to the angular frequency $\omega$ of the wave:
\begin{equation}
\label{local_wave}
\mathcal{O}\left(\frac{1}{t} h_{\mu \nu}\right) \ll \mathcal{O}(\omega h_{\mu \nu}).
\end{equation}
The time derivative of any of the scale factors goes like $1/t$, so the first and third sets of terms of Eqs. (\ref{h22}) and (\ref{h23}) go like $h_{\mu \nu}/t^2$ and $\omega h_{\mu \nu}/t$ respectively and are negligible compared to the other terms, which are second derivatives of $h_{\mu \nu}$ and go like $\omega^2 h_{\mu \nu}$. Therefore, Eqs. (\ref{h22}) and (\ref{h23}) simplify to the same form:
\begin{equation}
\frac{1}{a^2}\left(\frac{d^2}{dx^2}+\frac{1}{(1+\beta)^2}\frac{d^2}{dy^2}+\frac{1}{(1+\gamma)^2}\frac{d^2}{dz^2}\right) \phi - \ddot \phi =0,
\end{equation}
where $\phi(\eta,x,y,z)$ stands in for $h_{22}(\eta,x,y,z)$ and $h_{32}(\eta,x,y,z)$.

We write the expression in terms of ``approximately conformal" time $\eta$, which we define as 
\begin{equation}
\frac{1}{dt} \equiv \frac{1}{a}\frac{1}{d\eta}.
\end{equation}
This results in
\begin{equation}
\frac{1}{a^2}\left(\frac{d^2}{dx^2}+\frac{1}{(1+\beta)^2}\frac{d^2}{dy^2}+\frac{1}{(1+\gamma)^2}\frac{d^2}{dz^2}\right) \phi + \frac{a'}{a^3} \phi'- \frac{1}{a^2} \phi'' = 0,
\end{equation}
where $'$ denotes differentiation with respect to $\eta$.  We write this equation as
\begin{equation}
\label{phiprime}
\frac{1}{a^2}\tilde{\nabla}^2 \phi +\frac{a'}{a^3} \phi' - \frac{1}{a^2} \phi''=0,
\end{equation}
where we have defined
\begin{equation}
\tilde{\nabla}^2 \equiv \frac{d^2}{dx^2}+\frac{1}{(1+\beta)^2}\frac{d^2}{dy^2}+\frac{1}{(1+\gamma)^2}\frac{d^2}{dz^2}.
\end{equation}
The second term in Eq. (\ref{phiprime}) goes like $\omega h_{\mu \nu}/t$ and, as before, is negligible.
Therefore, the equation to solve is now
\begin{equation}
\label{eq_solve}
\tilde{\nabla}^2 \phi - \phi'' = 0.
\end{equation}
We assume a solution of the form 
\begin{equation}
\label{phi}
\phi(\eta,x,y,z) = \frac{g(\eta - \tilde{r})}{\tilde{r} a(\eta)},
\end{equation}
where 
\begin{equation}
\label{rtilde}
\tilde{r}= \sqrt{x^2+(1+\beta)^2 y^2+(1+\gamma)^2 z^2}.
\end{equation}
This $\tilde{r}$ is very similar to the Cartesian $r=\sqrt{x^2+y^2+z^2}$ in the FLRW background.  The factors of $\beta$ and $\gamma$ which scale the scaling factor $a$ for the $y$ and $z$ directions can be thought of equivalently as rescalings of $y$ and $z$:
\begin{equation}
y \rightarrow (1+\beta) y, ~~ z \rightarrow (1+\gamma) z.
\end{equation}
Substituting Eq. (\ref{phi}) into Eq. (\ref{eq_solve}) results in 
\begin{align}
\tilde{\nabla}^2 \left( \frac{g}{\tilde{r} a} \right)  -\left(\frac{g''}{\tilde{r}a}  -\frac{g \tilde{r}''}{\tilde{r}^2a} +2\frac{g \tilde{r}'^2}{\tilde{r}^3a}+ \frac{g\tilde{r}'a'}{\tilde{r}^2a^2} -\frac{ga''}{\tilde{r}a^2}+\frac{ga'\tilde{r}'}{\tilde{r}^2a^2}+ 2\frac{ga'^2}{\tilde{r}a^3}\right)=0.
\end{align}
From Eq. (\ref{rtilde}), we see that $\tilde{r}'$ is non-zero only at or above the first-order level in $\dot \beta$ and $\dot \gamma$.  Within the second set of parentheses of the above equation, all terms go like $h_{\mu \nu}/t^2$ except for the first term, so we keep only that first term.
We now have the final form of the wave equation to solve:
\begin{equation}
\label{final_form}
\tilde{\nabla}^2 \left( \frac{g}{\tilde{r} a} \right) - \frac{g''}{\tilde{r}a} = 0.
\end{equation}
The left-hand side of Eq. (\ref{final_form}), to first order in $\beta$, $\gamma$, $\dot \beta$, and $\dot \gamma$, is
\begin{equation}
\label{remainder}
\frac{2y^2 \beta' g''(\eta-r)}{r^2 a} + \frac{2z^2 \gamma' g''(\eta-r)}{r^2 a} - \frac{y^2 \beta'' g'(\eta-r)}{r^2 a} - \frac{z^2 \gamma'' g'(\eta-r)}{r^2 a}.
\end{equation}
Given that $g=\tilde{r} a \phi$ (from Eq. (\ref{phi})), we see that $g' \sim \omega g$ for its biggest contributing term, and each term in Eq. (\ref{remainder}) is thus negligible under the local wave approximation since $\beta', ~\gamma' \sim \beta/t, ~\gamma/t$.  Therefore, the {\it ansatz} Eq. (\ref{phi}) satisfies Eq. (\ref{final_form}).  The solution for the corresponding GW equation that results when an FLRW background is used (see Eq. 4.182 of \cite{Maggiore}) is the same as the solution Eq. (\ref{phi}) when 
\begin{equation}
r \rightarrow \tilde{r}.
\end{equation}  
The solution for a generic binary system (black hole-black hole, neutron star-black hole, neutron star-neutron star, etc) in circular orbit to first order in $v/c$ can be made to include back-reaction with respect to an FLRW background in a simple way using the local wave approximation (Eqs. 4.191 and 4.192 from \cite{Maggiore}).  Because we have the same solution if we substitute $\tilde{r}$ for $r$, we can also express the solution for the two polarizations of GWs for our perturbative Bianchi I (PBI) background in a similar manner:
\begin{equation}
\label{hplusPBI}
h_{PBI,+} = \frac{G M_c(1+\mathrm{z})/c^2}{a(t_0)\tilde{r}(1+\mathrm{z})}\left(\frac{5G M_c (1+\mathrm{z})/c^3}{t_{coal}-t}\right)^{1/4}\cos{\left(-2\left(\frac{t_{coal}-t}{5G M_c(1+\mathrm{z})/c^3}\right)^{5/8}+\Phi_0\right)}\frac{1+\cos^2\iota}{2},
\end{equation}
\begin{equation}
\label{htimesPBI}
h_{PBI,\times}=\frac{G M_c(1+\mathrm{z})/c^2}{a(t_0)\tilde{r}(1+\mathrm{z})}\left(\frac{5G M_c (1+\mathrm{z})/c^3}{t_{coal}-t}\right)^{1/4}\sin{\left(-2\left(\frac{t_{coal}-t}{5G M_c(1+\mathrm{z})/c^3}\right)^{5/8}+\Phi_0\right)}\cos\iota,
\end{equation}  
where $t_0$ is the present time, $M_c$ is the chirp mass (a combination of the reduced mass and the total mass of the binary system), $\mathrm{z}$ is the redshift of the system, $t_{coal}$ is the coalescence time, $\Phi_0$ is an integration constant specifying the phase of the wave, and $\iota$ is the angle of inclination of the orbital plane of the system.  Only the first factor of these equations will be relevant for this work as the luminosity distance only depends on the first factor.  The cosine function containing $\eta_{obs}$ is the oscillatory part, while the the other factors determines the evolution of the envelope of the oscillation.  The luminosity distance in FLRW space is 
\begin{equation}
d_{FLRW}(\mathrm{z}) = a(t_0) r (1+\mathrm{z}),
\end{equation}
so the solution in an FLRW background, for which we have $r$ instead of $\tilde{r}$, can be written as:
\begin{equation}
\label{hplusFLRW}
h_{FLRW,+} = \frac{G M_c(1+\mathrm{z})/c^2}{d_{FLRW}(\mathrm{z})}\left(\frac{5G M_c (1+\mathrm{z})/c^3}{t_{coal}-t}\right)^{1/4}\cos{\left(-2\left(\frac{t_{coal}-t}{5G M_c(1+\mathrm{z})/c^3}\right)^{5/8}+\Phi_0\right)}\frac{1+\cos^2\iota}{2},
\end{equation}
\begin{equation}
\label{htimesFLRW}
h_{FLRW,\times}=\frac{G M_c(1+\mathrm{z})/c^2}{d_{FLRW}(\mathrm{z})}\left(\frac{5G M_c (1+\mathrm{z})/c^3}{t_{coal}-t}\right)^{1/4}\sin{\left(-2\left(\frac{t_{coal}-t}{5G M_c(1+\mathrm{z})/c^3}\right)^{5/8}+\Phi_0\right)}\cos\iota.
\end{equation}  
We will revisit these expressions later, after we have derived the luminosity distance in our perturbative Bianchi I framework.


\section{Luminosity Distance}

The luminosity distance in this metric mimics the same form as in the FLRW metric:
\begin{equation}
\frac{\xi}{d_{PBI}(\mathrm{z})} \approx \frac{1}{a(\eta_0)\tilde{r} (1+\mathrm{z})}.
\end{equation}
Notice that $\tilde{r}$ plays the role of $r$, and there is a conversion factor $\xi$ which accounts for the fact that comoving distance and luminosity distance are altered relative to their FLRW counterparts. It is defined as 
\begin{equation}
\xi = \left( {\frac{d_{PBI}}{\tilde{r}}} \right) \bigg/ \left( {\frac{d_{FLRW}}{r}} \right).
\end{equation}
In addition to the perturbative quantities $\beta$ and $\gamma$ that parameterize the anisotropy, $\xi$ is also a function of angle, redshift, and matter density fraction. 
 A universe dominated by matter and a cosmological constant is assumed.  The full expression is, using the expression for $d_{PBI}$ from \cite{Schucker:2014wca}, 
\begin{align}
\xi=&\bigg[\bigg(
    1 + 
    \frac{1}{3 (\sqrt{g (\mathrm{z} + 1)^3 + 1} - \sqrt{g + 1})} 
    \bigg(
        4 \left( \frac{\mathrm{z}}{I} - \sqrt{g + 1} \right) +
        \left(
            \frac{2 (\mathrm{z} + 1)}{I \sqrt{g (\mathrm{z} + 1)^3 + 1}} - 3
        \right) \times \nonumber \\
        &\left( \sqrt{g (\mathrm{z} + 1)^3 + 1} - \sqrt{g + 1} \right)
        \left(
            (\beta - 2 \gamma) (\cos^2{\theta} - \sin^2{\phi} \sin^2{\theta}) +
            (\beta + \gamma) \cos{2 \phi} \sin^2{\theta} 
        \right) 
    \bigg)
\bigg)^{1/2} \times \nonumber \\
&\bigg( 
1 - \frac{\sqrt{g (\mathrm{z} + 1)^3 + 1}}{
    3 (\sqrt{g (\mathrm{z} + 1)^3 + 1} - \sqrt{g + 1}) \sqrt{g + 1}
} 
\bigg(
    (\sqrt{g + 1} - \frac{\mathrm{z}}{I})(\beta + \gamma)\; + \nonumber \\
    &\frac{\mathrm{z} + 1}{I}
    \left( 
        1 - \frac{\sqrt{g + 1}}{\sqrt{g (\mathrm{z} + 1)^3 + 1}}
    \right)
    \left(
       \beta (1 - 3 \sin^2{\phi} \sin^2{\theta})+ 
       \gamma (1 - 3 \cos^2{\theta}) \right)
\bigg)
\bigg)
\bigg]^{-1}.
\end{align}

$I$ is the elliptic integral
\begin{equation}
I = -\int_{0}^{\mathrm{z}} \! \frac{1}{\sqrt{g(\mathrm{z}'+1)^3+1}} \, \mathrm{d}\mathrm{z}',
\end{equation}
$g$ relates to the the matter density fraction as
\begin{equation}
g = \frac{1}{\frac{1}{\Omega_m}-1},
\end{equation}
and $\theta$ and $\phi$ denote angular dependence. $\theta$ is the angle between the source and the $z$ axis, such that a source directly along $+z$ has $\theta=0$ and along $-z$ has $\theta=\pi$. $\phi$ is the azimuthal angle of the source, ranging from $0$ to $2\pi$ such that a source in the $x$ direction has $\phi=0$. In the next section we will choose a coordinate system to define these axes on the sky. \\
Now we can express Eqs. (\ref{hplusPBI}) and (\ref{htimesPBI}) as
\begin{equation}
\label{xihplus}
h_{PBI,+} = \xi \frac{G M_c(1+\mathrm{z})/c^2}{d_{PBI}(\mathrm{z})}\left(\frac{5G M_c (1+\mathrm{z})/c^3}{t_{coal}-t}\right)^{1/4}\cos{\left(-2\left(\frac{t_{coal}-t}{5G M_c(1+\mathrm{z})/c^3}\right)^{5/8}+\Phi_0\right)}\frac{1+\cos^2\iota}{2},
\end{equation}
\begin{equation}
\label{xihcross}
h_{PBI,\times}=\xi \frac{G M_c(1+\mathrm{z})/c^2}{d_{PBI}(\mathrm{z})}\left(\frac{5G M_c (1+\mathrm{z})/c^3}{t_{coal}-t}\right)^{1/4}\sin{\left(-2\left(\frac{t_{coal}-t}{5G M_c(1+\mathrm{z})/c^3}\right)^{5/8}+\Phi_0\right)}\cos\iota,
\end{equation}  
Thus, if one assumes an FLRW background when fitting a gravitational wave, and if our universe is more accurately described with a PBI background, the inference to the fitted luminosity distance would be off by a factor of $\xi$ according to Eqs. (\ref{hplusFLRW}) and (\ref{htimesFLRW}), which in turn would bias the inference of the Hubble parameter.  This difference in the waveform would manifest as a difference in the size of the envelope surrounding the oscillatory wave (see Fig. 1). 
\begin{figure}[h]
\begin{centering}
\includegraphics[width=11cm]{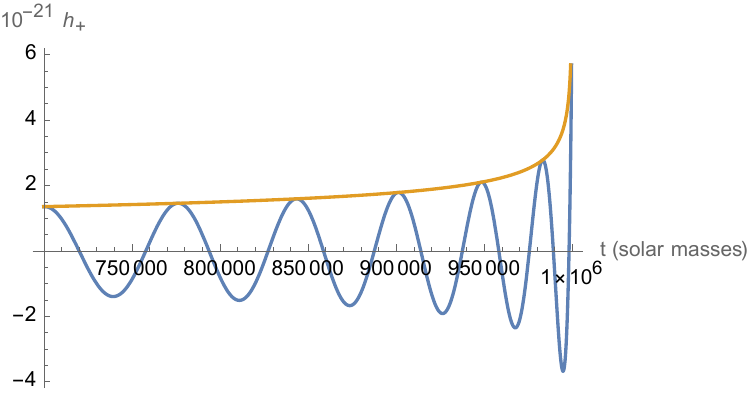}
\caption{Example gravitational waveform with the envelope indicated. We plot Eq. (\ref{xihplus}), and we use $G=c=1$, $M_c=65 M_\odot$, $t_{coal}=6.41 \times 10^6 M_\odot$ ($10^{-6}$ years) after $t=0$, $\mathrm{z}=6.8$, $\Phi_0=0$, $\iota=0$, $\xi=0.979$ (its angular-average value, from Eq. (\ref{avgxi})), and $d_{PBI}=1.6\times10^{22} M_\odot$ (17 billion light years).  For the same luminosity distance as for an FLRW background, the envelope would be affected by a factor of $\xi$.  This is true for the case of Eq. (\ref{xihcross}) as well, which would simply be phase-shifted relative to this plot with these chosen parameters.}
\end{centering}
\end{figure}

\section{Discussion and Results}
\begin{figure}[h]
\begin{centering}
\includegraphics[width=16cm]{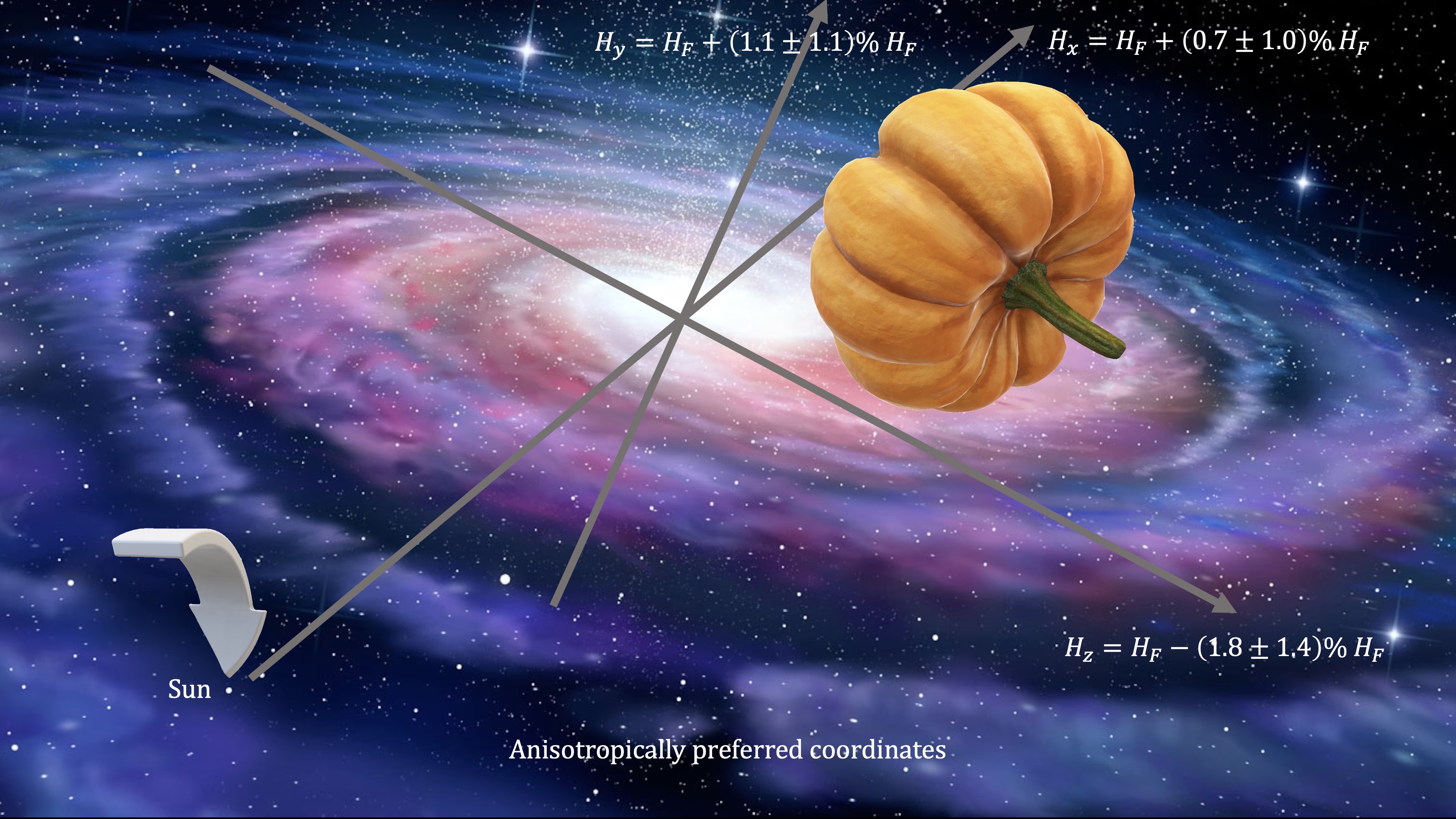}
\caption{Anisotropically preferred axes plotted against the Milky Way. The white arrow designates the location of the Sun.}
\label{pumpkin}
\end{centering}
\end{figure}

Recent supernova data constrain the level of anisotropy in the Bianchi I model \cite{Schucker:2014wca} to be
\begin{equation}
\beta= -0.69\%,
\end{equation}
\begin{equation}
\gamma= 4.29\%.
\end{equation}
Applying these constraints we obtain an angularly averaged $\xi$ value of 
\begin{equation}
\label{avgxi}
\bar{\xi} = 0.979.
\end{equation}
This result means that if we live in a PBI universe, inferences of luminosity distance are on average too large, manifesting in gravitational wave amplitudes which are overestimated by $2.1\%$. \\
Notice that these numerical constraints on anisotropy have at last forced us to choose a coordinate system. The axes have been established as follows:
\begin{itemize}
\item The direction of greatest anisotropy is $+z$.
\item In the plane orthogonal to that axis, the direction of greatest anisotropy is $+x$.
\item $+y$ is now uniquely determined by using a right-handed orthogonal basis.
\end{itemize}
Perhaps counterintuitively, the values of $\beta$ and $\eta$ do not directly correspond to the anisotropy in the $y$ and $z$ directions. To understand why, recall that the $x$ component of the perturbative Bianchi I metric is not perturbed. Also note that the average value of the Hubble constant must match that of an FLRW universe. Taking these two facts in combination, we see that $a(t)$, the scale factor along the $x$ axis, must take a different value in a PBI universe than in an FLRW one. 
The anisotropic stand-in for the FLRW $a(t)$ is defined as \cite{Schucker:2014wca}
\begin{equation} 
W(a) = \left(\frac{A^2}{a^2} + \frac{B^2}{b(a)^2} + \frac{C^2}{c(a)^2}\right)^{-1/2}.
\end{equation}
where $A$, $B$, and $C$ are integration constants, and redshift in the Bianchi I metric is approximately 
\begin{equation} 
\mathrm{z} \approx \frac{W(a(t_0))}{W(a(t_i))}-1,
\end{equation}
So $H_F$ (the isotropic Friedmann Hubble constant) is $\frac{\dot{W}}{W}$ rather than $\frac{\dot{a}}{a}$. It is useful to define a Hubble constant for each direction
\begin{align} 
H_x &=\frac{\dot{a}}{a} \\
H_y &=\frac{\dot{b}}{b} \\
H_z &=\frac{\dot{c}}{c} 
\end{align}
with the constraint that 
\begin{equation} 
H_F = \frac{H_x+H_y+H_z}{3}.
\end{equation}
Solving for $\beta$ and $\eta$ in terms of these Hubble parameters reveals the $x$ direction ``mixing" that we intuitively argued must exist.
\begin{align} 
\beta &\propto H_y-H_x \\
\gamma &\propto H_z-H_x 
\end{align}
Fig. (\ref{pumpkin}) shows how these preferred axes related to the Milky Way. If a true anisotropy exists, our universe would be overall ``pumpkin shaped" in the orientation shown.

\begin{figure}[h]
\begin{centering}
\includegraphics[width=17cm]{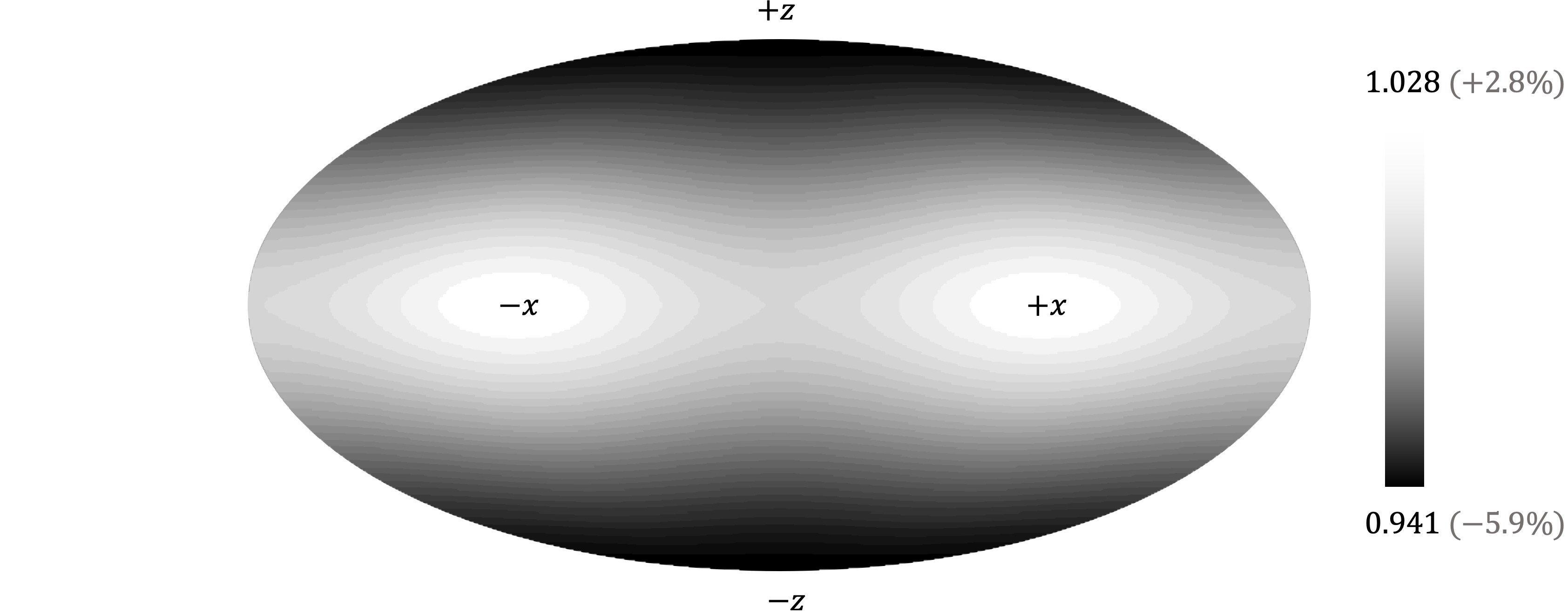}
\caption{Angular dependence of $\xi$.}
\label{ang_dep}
\end{centering}
\end{figure}

The $z$ axis lies almost exactly in the galactic plane and is nearly orthogonal to the Sun-galactic center line. The $x$ axis tips slightly out of the plane \cite{Schucker:2014wca}. After choosing these axes, thereby concretizing the angles from our $\xi$ expression, we plot the variation of $\xi$ over the sky in Fig. (\ref{ang_dep}). 

Sources along the $x$ axis would lead to the greatest possible overestimation of luminosity distance, up to 2.8\%, and sources along the $z$ axis would have the largest possible underestimation of luminosity distance, up to 5.9\% lower than what would be expected in the FLRW case.  

\section{Conclusion}
In this work, we assumed a perturbative Bianchi I background which was conducive to direct comparison with the FLRW GW solution for a binary system in circular orbit to first order in $v/c$ which includes back-reaction with respect to an FLRW background ("local wave" approximation).  We formulated the luminosity distance in this framework and found that inferring the luminosity distance from a GW event may result in error if the wrong background spacetime is assumed for the model fit, as much as up to 5.9\%.

Assuming an incorrect background spacetime through which GWs propagate to our observatories has a very small effect on the waveform in general, but the cumulative effect of inferring a slightly biased luminosity distance for several observations may add up to a bad inference of the Hubble parameter, one that is significantly erroneous.  Though the typical luminosity distance uncertainty for current detectors such as LIGO, VIRGO, and KAGRA is 30-60\% \cite{061102, 1811.12907, Chen2018}, future detectors should be able to probe the discrepancy we have derived in this work.  Current forecasts for future detectors' uncertainty in luminosity distance estimation are $\lesssim 10$\% for redshifts $\mathrm{z} \lesssim5$ for the Einstein Telescope and Cosmic Explorer \cite{1912.02622, 1907.04833, 1206.0331}, and 0.1-10\% for the space-based LISA, depending on the source \cite{1511.05581, 0906.3752, 1601.07112}.  Therefore, assuming the wrong background metric will have a measurable effect for future gravitational wave detectors.



\end{document}